\newcommand*{\ATLASLATEXPATH}{}
\documentclass[cernpreprint,texlive=2016,UKenglish]{\ATLASLATEXPATH atlasdoc}
 
\usepackage{\ATLASLATEXPATH atlaspackage}
\usepackage{\ATLASLATEXPATH atlasheavyion}
\usepackage{lipsum}
\usepackage{todonotes}
 
\newcommand{\Nobs}{59\xspace}
\newcommand{\Nbkg}{12\xspace}
\newcommand{\dNbkg}{3\xspace}
\newcommand{\NsigExp}{30\xspace}
\newcommand{\CFactor}{0.350\xspace}
\newcommand{\dCFactor}{0.024\xspace}
\newcommand{\SigmaMeas}{\SI[parse-numbers=false]{78 \pm 13~\textrm{(stat.)} \pm 7~\textrm{(syst.)} \pm 3~\textrm{(lumi.)}}{\nb}\xspace}
\usepackage[backref=false]{\ATLASLATEXPATH atlasbiblatex}
 
\usepackage{\ATLASLATEXPATH atlascontribute}
 
\usepackage{\ATLASLATEXPATH atlasphysics}

\graphicspath{{logos/}{figures/}}
 
\usepackage{ANA-HION-2018-19-PAPER-defs}

\AtlasTitle{Observation of light-by-light scattering in ultraperipheral Pb+Pb collisions with the ATLAS detector}

\AtlasAbstract{
This letter describes the observation of the light-by-light scattering process, $\gamma\gamma\to\gamma\gamma$, in \PbPb\ collisions at \sqrtNN\,=\,\SI{5.02}{\TeV}.
The analysis is conducted using a data sample corresponding to an integrated luminosity of \SI{1.73}{\inb}, collected in November 2018 by the ATLAS experiment at the LHC.
Light-by-light scattering candidates are selected in events with two photons produced exclusively, each with transverse energy $E_{\textrm{T}}^{\gamma} > \SI{3}{\GeV}$ and pseudorapidity $|\eta_{\gamma}| < 2.4$, diphoton invariant mass above $\SI{6}{\GeV}$, and small diphoton transverse momentum and acoplanarity.
After applying all selection criteria, \Nobs candidate events are observed for a background expectation of $\Nbkg\pm\dNbkg$ events.
The observed excess of events over the expected background has a significance of 8.2 standard
deviations.
The measured fiducial cross section is \SigmaMeas.
}
 
\author{The ATLAS Collaboration}

\AtlasJournalRef{Phys. Rev. Lett. 123 (2019) 052001}
\AtlasDOI{10.1103/PhysRevLett.123.052001}
 
\PreprintIdNumber{CERN-EP-2019-051}
 
\hypersetup{pdftitle={ATLAS document},pdfauthor={The ATLAS Collaboration}}
\usepackage{textpos}
\clearpage
 
\begin{document}
 
\maketitle
 
\clearpage
\newpage

Light-by-light scattering, $\gamma\gamma\rightarrow\gamma\gamma$, is a quantum-mechanical process that is forbidden in the classical theory of electrodynamics~\cite{Euler:1935zz, Heisenberg1936}.
In the Standard Model (SM), the $\gamma\gamma\rightarrow \gamma\gamma$ reaction proceeds at one-loop level at order
$\alpha_{\textsc{em}}^4$ (where $\alpha_{\textsc{em}}$ is the fine-structure constant) via virtual
box diagrams involving electrically charged fermions (leptons and quarks) or $W^{\pm}$ bosons.
However, in various extensions of the SM, extra contributions are possible, making the measurement of $\gamma\gamma\rightarrow\gamma\gamma$ scattering sensitive to new physics.
Relevant examples are magnetic monopoles~\cite{Ginzburg:1998vb}, vector-like fermions~\cite{Fichet:2014uka} and axion-like particles~\cite{Knapen:2016moh, Bauer:2017ris}.
The light-by-light cross section is also sensitive to the effect of possible non-SM operators in an effective field theory~\cite{Ellis:2017edi,Akmansoy:2018xvd,Kostelecky:2018yfa}.
Light-by-light scattering graphs with electron loops also contribute to the anomalous magnetic moment of the electron and muon~\cite{Laporta:1992pa, Jegerlehner:2009ry}.
 
Strong evidence for this process in relativistic heavy-ion (\PbPb) collisions at the Large Hadron Collider~(LHC) has been reported by the ATLAS~\cite{Aaboud:2017bwk} and CMS~\cite{Sirunyan:2018fhl} collaborations with observed significances of 4.4 and 4.1 standard deviations, respectively.
Exclusive light-by-light scattering can occur in these collisions at impact parameters larger than about twice the radius of the ions, as demonstrated for the first time in Ref.~\cite{Enterria:2013yra}.
The strong interaction becomes less significant and the electromagnetic~(EM) interaction becomes more important in these ultraperipheral collision~(UPC) events.
In general, this allows to study processes involving nuclear photoexcitation, photoproduction of hadrons, and two-photon interactions~\cite{Bertulani:2005ru, Baltz:2007kq}.
The EM fields produced by the colliding Pb nuclei can be described as a beam of quasi-real photons with a small virtuality of $Q^2\,<\,1/R^2$, where $R$ is the radius of the charge distribution and so $Q^2\,<\,\SI{e-3}{\GeV\squared}$~\cite{Fermi:1925fq,PhysRev.45.729}.
The cross section for the elastic reaction $\textrm{Pb+Pb}\,(\gamma\gamma)\rightarrow \textrm{Pb}\textrm{+}\textrm{Pb}\,\gamma\gamma$ can then be calculated by convolving the appropriate photon flux with the elementary cross section for the process $\gamma\gamma\rightarrow \gamma\gamma$.
Since the photon flux associated with each nucleus scales with the square of the number of protons, the cross section is strongly enhanced relative to proton--proton ($pp$) collisions.
 
The $\gamma\gamma\rightarrow \gamma\gamma$ reaction has also been measured in photon scattering in the Coulomb field of a nucleus (Delbrück scattering)~\cite{Wilson:1953zz,PhysRevD.8.3813,Schumacher:1975kv,PhysRevC.58.2844} and in the photon-splitting process~\cite{Akhmadaliev:2001ik}. A related process, in which initial photons fuse to form a pseudoscalar meson that subsequently decays into a pair of photons, has been studied at electron--positron colliders~\cite{Bartel:1985zw, Aihara:1985uc, Williams:1988sg,Babusci:2012ik}.

The previous ATLAS and CMS measurements were based on the
\PbPb\ dataset of $~\SI{0.4}{\inb}$ recorded in 2015 at a center-of-mass energy of \sqrtNN\,=\,\SI{5.02}{\TeV}~\cite{Aaboud:2017bwk,Sirunyan:2018fhl}.
The present Letter describes a new measurement exploiting  \SI{1.73}{\inb} of \PbPb collisions at \sqrtNN\,=\,\SI{5.02}{\TeV}, recorded in November 2018 with the ATLAS detector at the LHC.
The analysis follows the approach originally proposed in Ref.~\cite{Enterria:2013yra}, which was the basis of the initial ATLAS measurement.

The ATLAS detector~\cite{ATLAS_detector_paper} is a multipurpose particle detector that covers
nearly the entire solid angle around the interaction point (IP)~\cite{atlas_footnote}.
It consists of an inner detector~(ID) for charged-particle tracking in the pseudorapidity region $|\eta|$\,<\,2.5, EM and hadronic calorimeters that provide energy measurements up to $|\eta|$\,=\,4.9, and a muon spectrometer that covers $|\eta|$\,<\,2.7.
Forward calorimeters (FCal) cover the range of 3.2\,<\,$|\eta|$\,<\,4.9.
The zero-degree calorimeters~(ZDC), located along the beam axis at 140\,m
from the IP on both sides, detect neutral particles, including neutrons emitted from the nucleus.
 
The final-state signature of interest is the exclusive production of two photons, $\textrm{Pb+Pb}\,(\gamma\gamma)\rightarrow \textrm{Pb}^{(\ast)}\textrm{+}\textrm{Pb}^{(\ast)}\,\gamma\gamma$, where the diphoton final-state is measured in the central detector, and the incoming Pb ions survive the EM interaction, with a possible EM excitation~\cite{ALICE:2012aa}, denoted by $(^{\ast})$.
Hence, the final state consists of two low-energy photons and no further activity in the detector, and in particular no reconstructed charged-particle tracks originating from the IP.
 
A two-level trigger system was used to select events online~\cite{Aaboud:2016leb}. It consists of a level-1 trigger implemented using a combination of custom electronics and programmable logic, and a software-based high-level trigger~(HLT).
Candidate diphoton events were recorded using a dedicated trigger for events with moderate activity in the calorimeter but little additional activity in the detector.
At level-1 a logical OR of two conditions was required: at least one EM cluster with $\et\,>\,\SI{1}{\GeV}$ in coincidence with a total \et\ of \SIrange{4}{200}{\GeV}
measured in the calorimeter, or at least two
EM clusters with $\et\,>\,\SI{1}{\GeV}$ with total \et\ measured in the calorimeter below \SI{50}{\GeV}. The upper bound on the total \et\ was optimized to be fully efficient for signal events, while allowing the rejection of events from nonperipheral \PbPb collisions.
At the HLT, the total FCal \et on each side of the IP was required to be consistent with noise (FCal veto), and  the number of hits in the pixel detector (part of the ID) was required to be at most 15.

Simulated $\gamma\gamma\rightarrow\gamma\gamma$ signal events were generated using the SuperChic~3.0 Monte Carlo~(MC) generator~\cite{Harland-Lang:2018iur}. This program takes into account box diagrams with charged leptons, quarks and $W^{\pm}$ bosons.
An alternative signal sample was generated using calculations from Ref.~\cite{Klusek-Gawenda:2016euz}.
These calculations were then folded with the Pb+Pb photon flux taken from the Starlight 2.0 MC generator~\cite{Klein:2016yzr}. The theoretical uncertainty of the cross section is mainly due to the limited knowledge of the nuclear form factors and initial photon fluxes.
This is extensively studied in Refs.~\cite{Sirunyan:2018fhl, Azevedo:2019fyz}, and the relevant uncertainty is estimated to be $10\%$ within the fiducial phase space of the measurement.
Higher-order corrections, which are not included in the calculations, are also part of the
theoretical uncertainty  and amount to 1--3\% in the fiducial phase space~\cite{Bern:2001dg, Klusek-Gawenda:2016nuo}.
 
The exclusive diphoton final state can also be produced via the strong interaction through a quark loop in the exchange of
two gluons in a color-singlet state. This central exclusive production~(CEP) process, $gg\rightarrow\gamma\gamma$,
was also modeled using SuperChic~3.0.
This process has a large theoretical uncertainty, of $\mathcal{O}(100\%)$~\cite{Aaltonen:2011hi}, hence the absolute normalization of this background is determined using a control region in the data, as explained later.
The \ggee\ process is a potential background when both leptons are reconstructed as photons, but is also used for calibration studies in the analysis.
The process was modeled with the Starlight 2.0 generator. Its production cross section is computed by combining
the Pb+Pb photon flux with the leading-order formula for \ggee.
Two-photon production of quark--antiquark pairs, with their subsequent decay into multiple hadrons, was modeled using \textsc{Herwig}++ 2.7.1~\cite{Bahr:2008pv},
where the initial photon fluxes from $pp$ collisions are implemented.
The sample was then normalized to cover the differences in the photon fluxes between \PbPb\ and $pp$
collisions.
All simulated events make use of a detector simulation~\cite{Aad:2010ah} based on GEANT4~\cite{Agostinelli:2002hh} and are reconstructed with the standard ATLAS reconstruction software.

Photons are reconstructed from EM clusters in the calorimeter~\cite{ATL-PHYS-PUB-2017-022} and tracking information provided by the ID, which allows to identify photon conversions~\cite{Aaboud:2018yqu}.
An energy calibration specifically optimized for photons~\cite{Aaboud:2018ugz} is applied to account for energy loss before the calorimeter and both lateral and longitudinal shower leakage.
Photons in MC samples are corrected~\cite{Aaboud:2018yqu} for known mismodeling of quantities that describe the properties (``shapes'') of the associated EM showers.
 
The photon particle identification~(PID) in this analysis is based on a selection of these shower-shape variables, optimized for the signal events. Only photons with $\et\,>\,\SI{3}{\GeV}$ and $|\eta|\,<\,2.37$, excluding the calorimeter transition region $1.37\,<\,|\eta|\,<\,1.52$, are considered.
This allows for good separation between prompt photons and fake signatures due to calorimeter noise, cosmic-ray muons, or nonprompt photons originating from the decay of neutral hadrons.
The photon PID is based on a neural network trained on background photons extracted from data and on photons from the signal MC sample. The selection of background photons follows the procedure established in Ref.~\cite{Aaboud:2017bwk}.
 
Selected events are required to have exactly two photons satisfying the above selection criteria,
with a diphoton invariant mass ($\Minvgg$) greater than \SI{6}{\GeV}.
In order to suppress the \ggee\ background, events are rejected if they have a charged-particle track with
$\pt > \SI{100}{\MeV}$, $|\eta|\,<\,2.5$, and at least six hits in the pixel and microstrip
detectors, including at least one pixel hit.
To further suppress \ggee\ events with poorly reconstructed charged-particle tracks, candidate events are required to have no ``pixel tracks''
matched to  a photon candidate within $|\Delta\eta|\,<\,0.5$.
Pixel tracks are reconstructed using information from the pixel detector only.
They are required to have $\pt > \SI{50}{\MeV}$, $|\eta|\,<\,2.5$, and at least three hits in the pixel detector.
According to the MC simulation, these requirements reduce the fake photon background from the dielectron final state by a factor of $10^4$, while being 93\% efficient for \gggg\ signal events.
 
To reduce other fake-photon backgrounds, such as cosmic-ray muons, the transverse momentum of the diphoton system~(\ptgg) is required to be below \SI{1}{\GeV} for $\Minvgg\,<\,\SI{12}{\GeV}$ and below \SI{2}{\GeV} for $\Minvgg\,>\,\SI{12}{\GeV}$. To reduce prompt-photon background from CEP $gg\rightarrow\gamma\gamma$ reactions, an additional requirement on the reduced acoplanarity, $\aco = (1-|\Delta\phi_{\gamma\gamma}|/\pi)<0.01$, is used,
which is expected to have $(86\pm1)\%$ selection efficiency for the signal.
This efficiency is estimated using simulated signal events, and the uncertainty is due to modeling of the photon angular resolution in simulation.
The above requirements define the fiducial region for the signal measurement.
 
Exclusive dielectron pairs from the reaction $\textrm{Pb+Pb}\,(\gamma\gamma)\rightarrow \textrm{Pb}^{(\ast)}\textrm{+}\textrm{Pb}^{(\ast)}\,e^+e^-$ are used for various aspects of the analysis, in particular to validate the EM calorimeter energy scale and resolution~\cite{Aaboud:2018ugz}. To select \ggee candidates, events are required to pass the same trigger as for the diphoton selection.
Each electron is reconstructed from an EM energy cluster in the calorimeter matched to a track in the ID~\cite{Aaboud:2019ynx}. The \ggee events are selected by requiring exactly two oppositely charged electrons, no further charged-particle tracks coming from the interaction region, and dielectron reduced acoplanarity, $\aco < 0.01$.
The observed \ggee event yield in data is compatible with that expected from simulation.

\begin{figure}[t!]
\begin{center}
\includegraphics[width=0.328\linewidth]{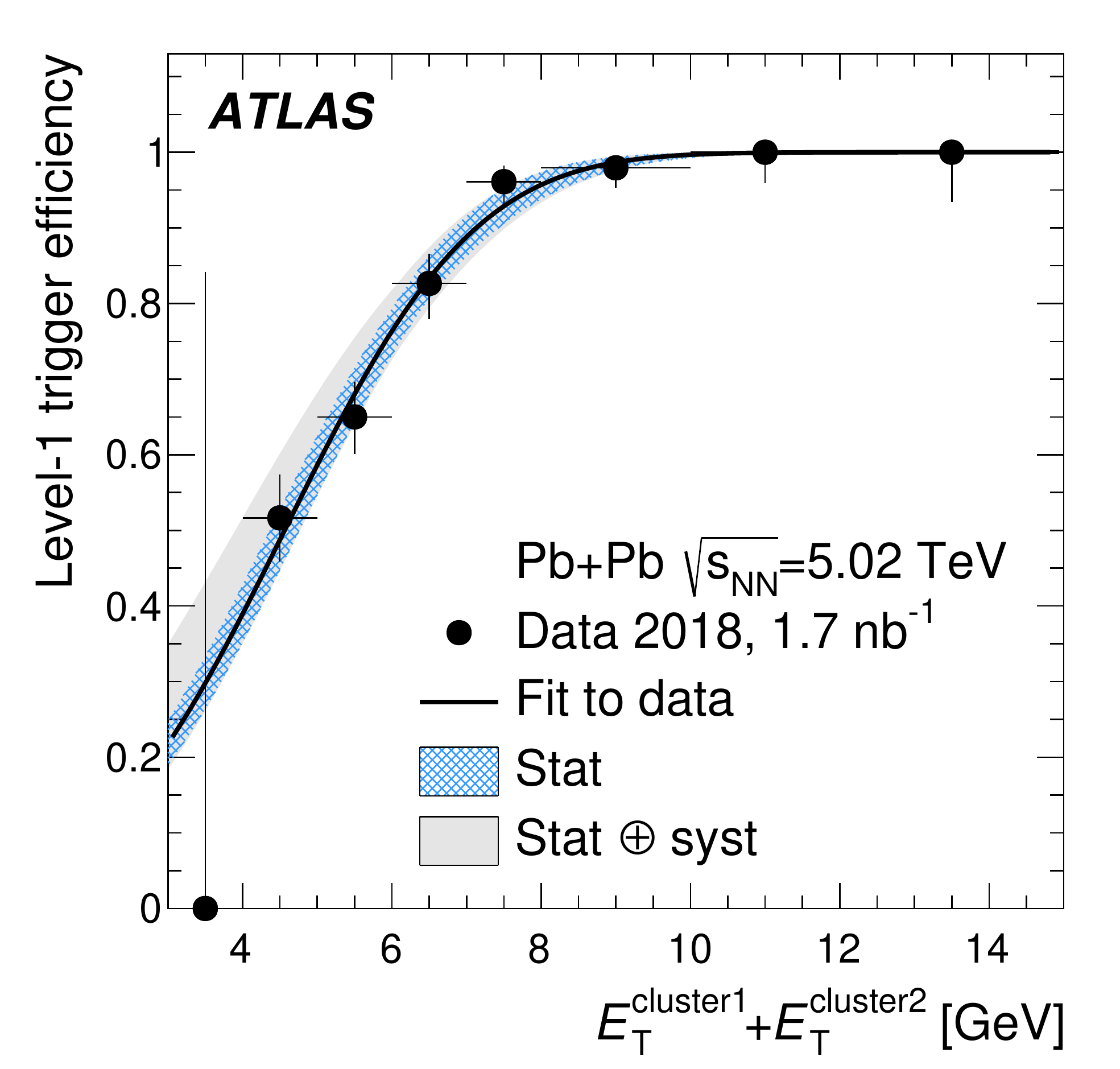}
\includegraphics[width=0.328\linewidth]{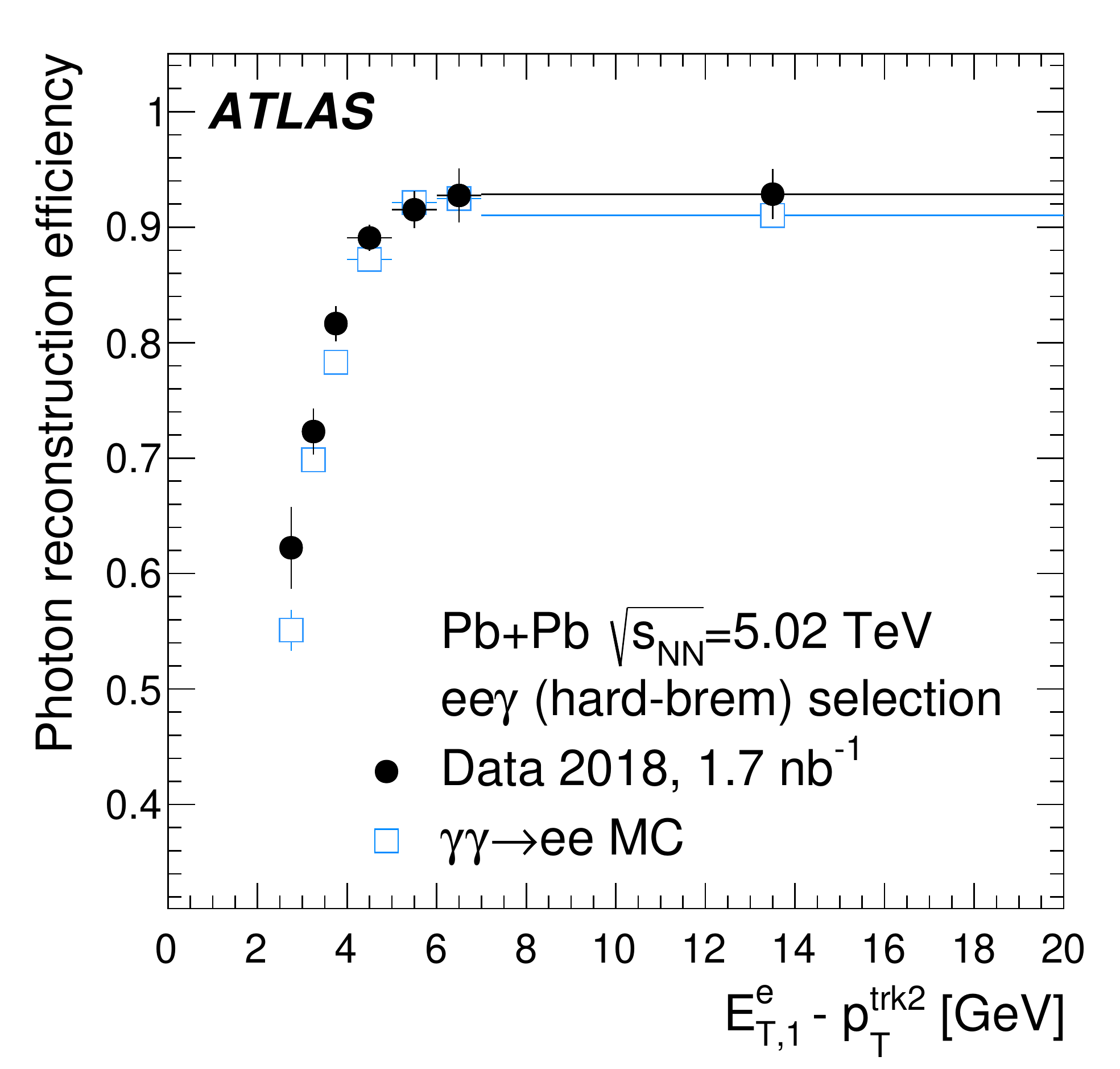}
\includegraphics[width=0.328\linewidth]{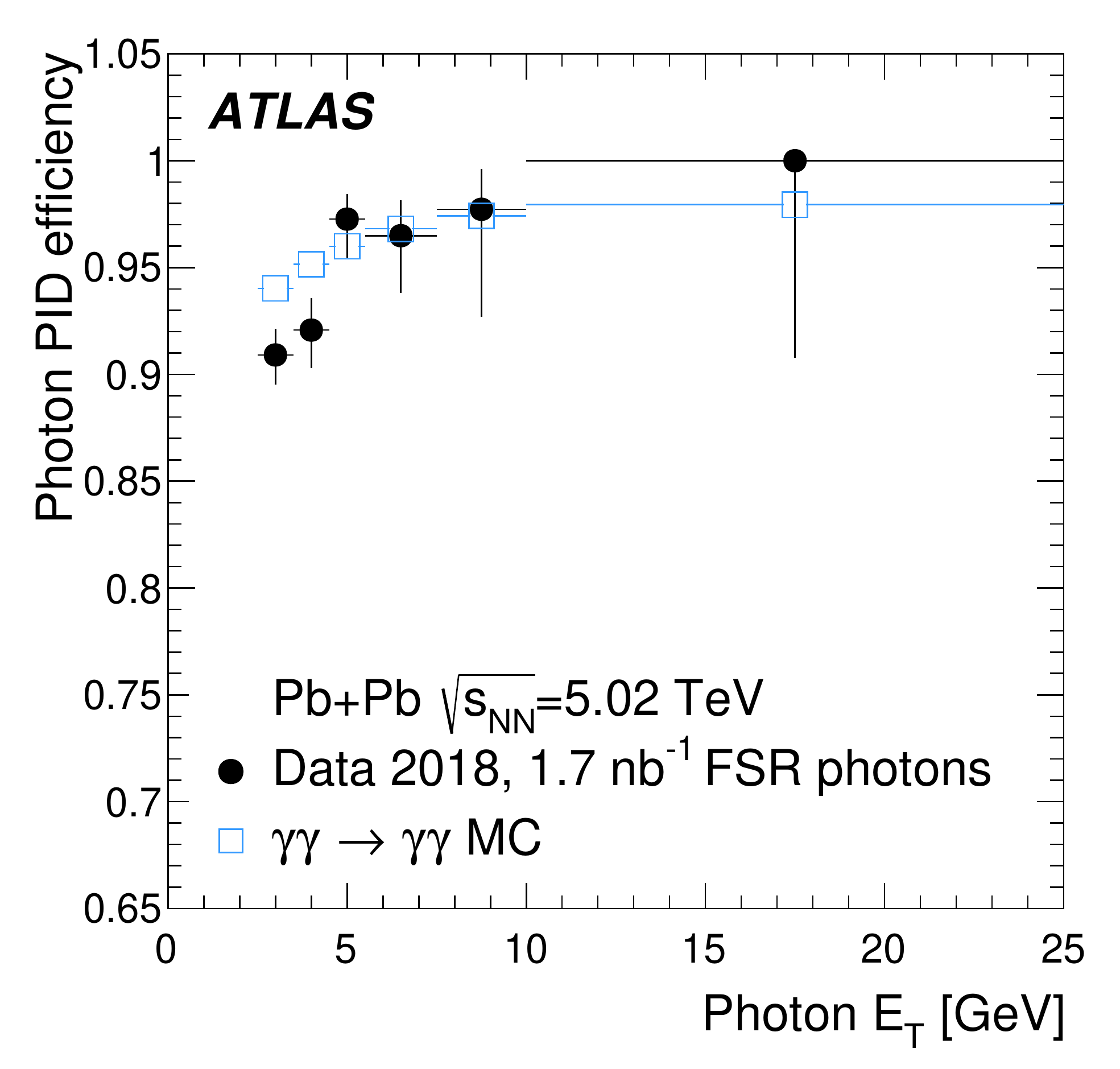}
\put(-375,-8){{\footnotesize(a)}}\put(-225,-8){{\footnotesize(b)}} \put(-75,-8){{\footnotesize(c)}}
\caption{(a) Measured level-1 trigger efficiency as a function of the reconstructed transverse energy in \ggee\ events, (b) photon reconstruction efficiency as a function of the photon \eT\ (approximated with $E^e_\mathrm{T,1}-\pT^\textrm{trk2}$, where trk2 denotes the track of the second leading electron), and (c) photon particle-identification efficiency as a function of the photon \et.}
\label{fig:LbLDetectorLevel}
\end{center}
\end{figure}

The level-1 trigger efficiency is estimated with \ggee\ events passing an independent trigger.
The level-1 trigger efficiency as a function of the electron EM cluster transverse energy sum, \etclone+\etcltwo, reaches $60\%$ at \SI{5}{\GeV} and $75\%$ at \SI{6}{\GeV}, with the fully efficient plateau reached at around \SI{10}{\GeV}, as shown in Figure~\ref{fig:LbLDetectorLevel}(a).
The measured efficiency is parameterized and used to correct the trigger response in the simulation.
To test the stability of the results, the analysis is repeated using tighter or looser dielectron event selection criteria, and the resulting differences are taken as a systematic uncertainty.
The FCal veto efficiency is estimated using \ggee\ events selected  with a dedicated control trigger without involving the FCal requirement. It is estimated to be $(99.1 \pm 0.6)\%$.
Due to the high hit-reconstruction efficiency and relatively low conversion probability of signal photons in the pixel detector, the inefficiency of the pixel veto requirement at the trigger level is found to be negligible.

The photon reconstruction efficiency is extracted from data using \ggee events, where one of the electrons emits a hard bremsstrahlung photon due to interaction with the material of the detector. The analysis is performed for events with exactly one identified electron and exactly two reconstructed charged-particle tracks, and a tag-and-probe method is used as described in Ref.~\cite{Aaboud:2017bwk}. The resulting photon reconstruction efficiency is shown in Figure~\ref{fig:LbLDetectorLevel}(b). It rises from about 60\% at $\eT\,=\,\SI{2.5}{\GeV}$ to 90\% at $\eT\,=\,\SI{6}{\GeV}$ and is used to derive simulation-to-data correction factors.
 
High-\pt\ exclusive dilepton production~(\ggll, where $\ell=e,~\mu$) with final-state radiation~(FSR) is used to measure the photon PID efficiency, defined as the probability for a reconstructed photon to satisfy the identification criteria. Events with exactly two oppositely charged tracks with $\pt\,>\,\SI{0.5}{\GeV}$ are selected from UPC triggered events. In addition, a requirement to reconstruct a photon candidate with $\et\,>\,\SI{2.5}{\GeV}$ and $|\eta|\,<\,1.37$ or $1.52\,<\,|\eta|\,<\,2.37$ is imposed. A photon candidate is required to be separated from each track by fulfilling $\Delta R\,>\,0.3$~\cite{atlas_footnote} to avoid leakage between the photon and the electron clusters. The FSR event candidates are required to have $\pTttg\,<\,\SI{1}{\GeV}$ requirement, where \pTttg\ is the transverse momentum of the three-body system consisting of  the two tracks and the photon candidate. Figure~\ref{fig:LbLDetectorLevel}(c) shows the photon PID efficiency as a function of reconstructed photon \et, where the measurement from data is compared with the one extracted from the signal MC sample.
Based on these studies, MC events are corrected using photon \et-dependent simulation-to-data correction factors.
The systematic uncertainty on the photon reconstruction and PID efficiencies is estimated by parameterizing the
correction factors as a function of the photon $\eta$ instead of the photon \et.
 
The two electrons exhibit balanced transverse momentum with an unbalance, $|\pt^{e^+}-\pt^{e^-}|$,
expected to be below \SI{30}{\MeV}. This is much smaller than the EM calorimeter energy resolution,
which can thus be measured by the difference $\etclone-\etcltwo$. Below \SI{10}{\GeV} electron $\et$
the relative energy resolution is found to be between 8\% and 10\% and is well reproduced by the MC
simulation. The EM energy scale is validated using the ratio of the electron cluster $\et^e$ to the
electron track $\pt^{\textrm{trk}}$.

The $\ggee$ process can be a source of fake diphoton events,
since misidentification of electrons as photons can occur when the electron track is not reconstructed or the electron emits a hard bremsstrahlung photon.
The $\ggee$ yield in the signal region is evaluated using a data-driven method. Two control regions (CRs) are defined with exactly two
photons passing the signal selection, but requiring also one or two associated pixel tracks.
The event yield observed in these two CRs is extrapolated to the signal region using the probability to miss the electron pixel track if the electron track is not reconstructed (\pmis).
It is measured in a region with exactly one charged-particle track and two photons with $\aco < 0.01$.
In order to verify the stability of the \pmis\ evaluation method, the $\aco$ requirement is dropped and the difference with the nominal selection is taken as a systematic uncertainty.
This leads to $\pmis\,=\,(47\pm9)\%$.
The number of \ggee events in the signal region is estimated to be $7\pm 1~\textrm{(stat.)} \pm 3~\textrm{(syst.)}$, where the uncertainty accounts for the CR statistical uncertainty, the $\pmis$ uncertainty, and the difference found between the two CRs.

\begin{figure}[t!]
\begin{center}
\includegraphics[width=0.328\linewidth]{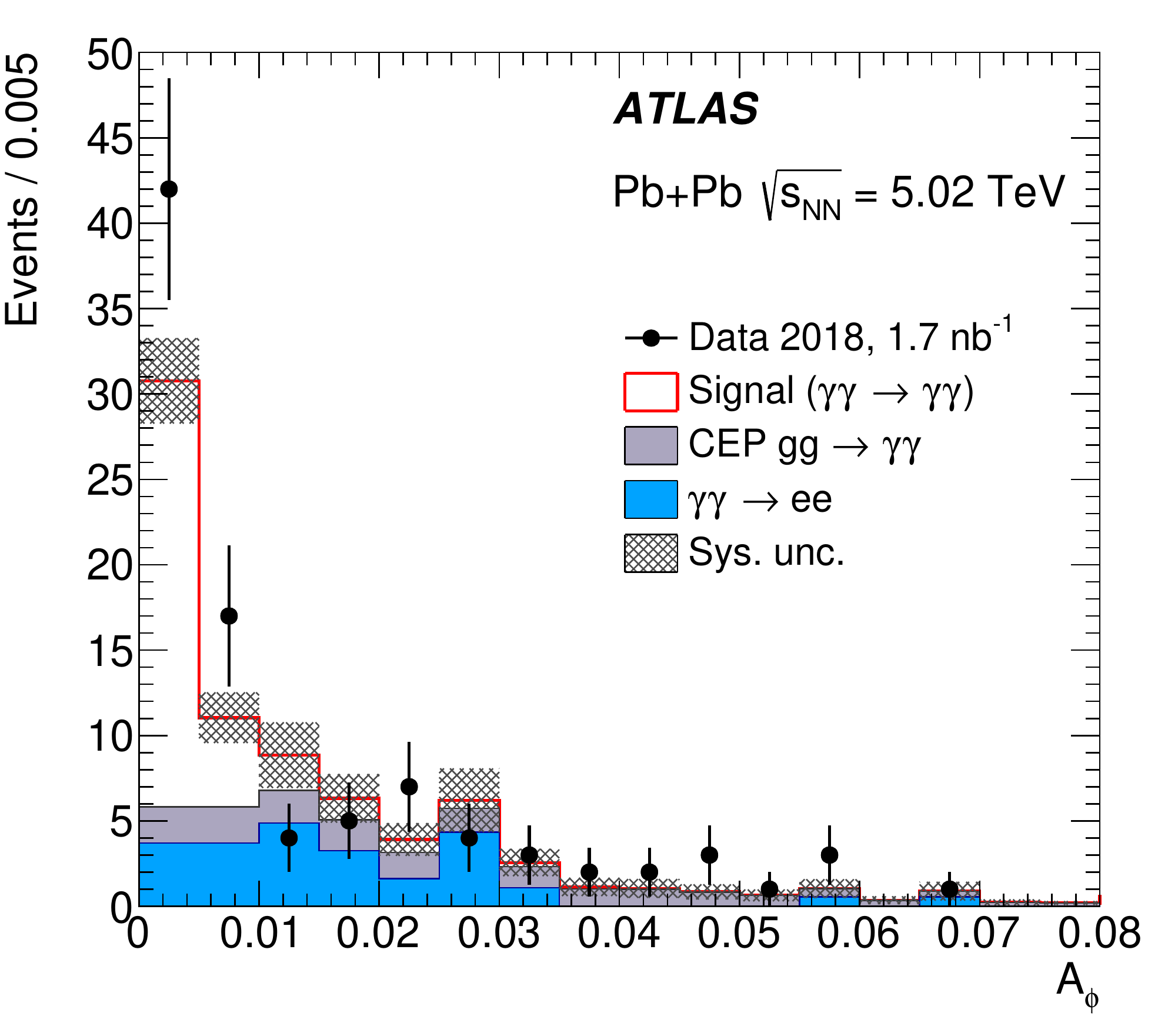}
\includegraphics[width=0.328\linewidth]{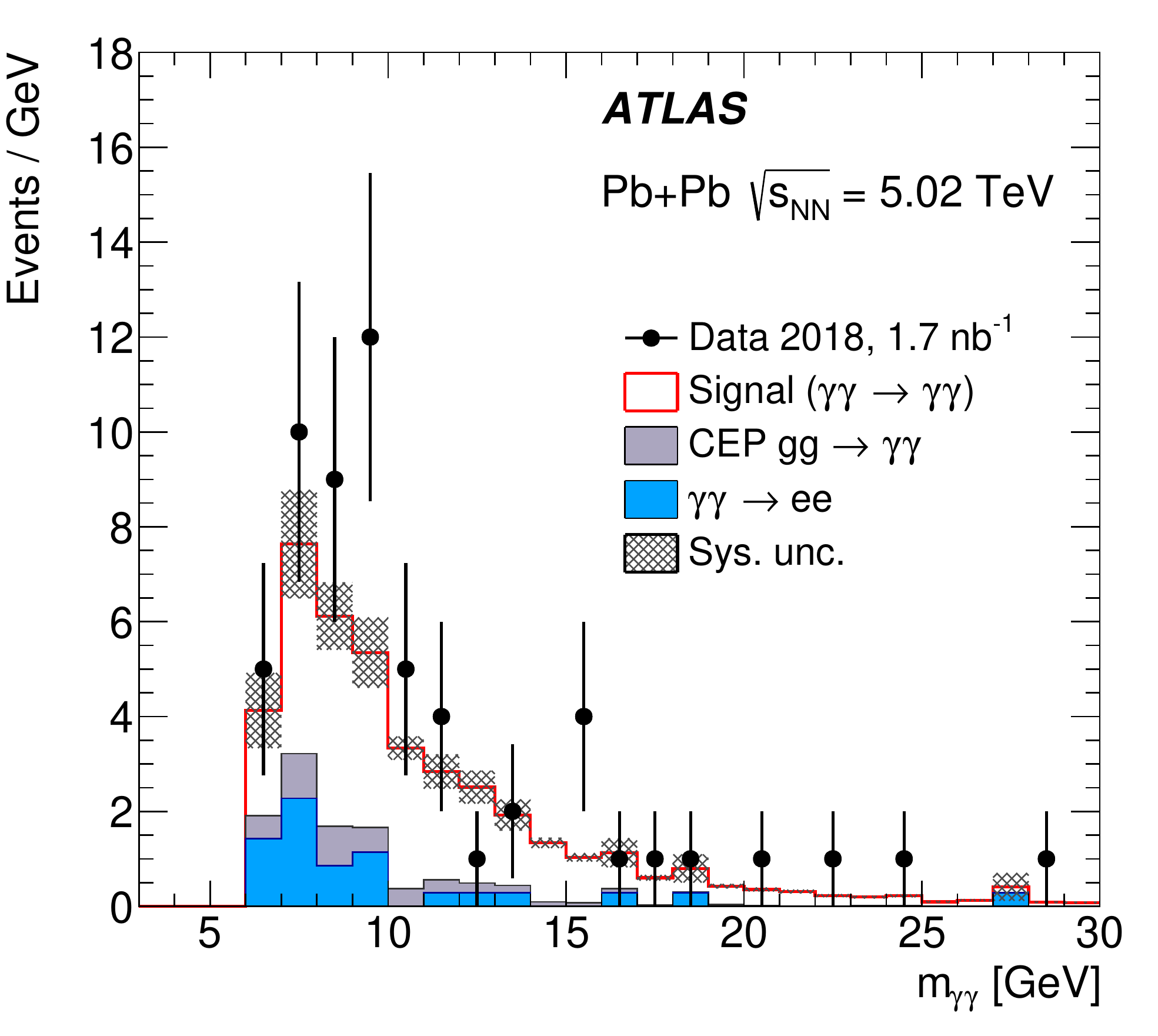}
\includegraphics[width=0.328\linewidth]{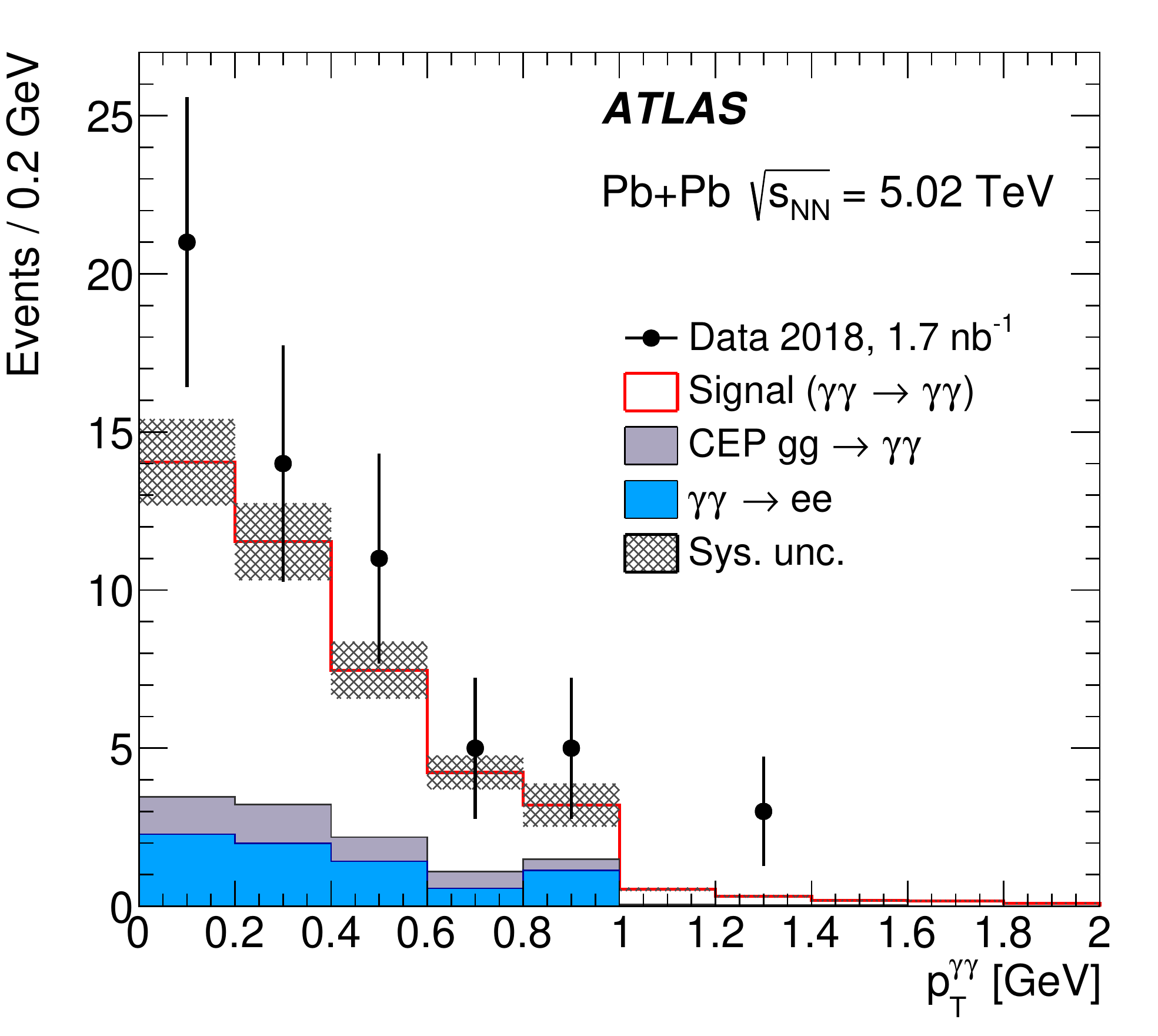}
\put(-375,-8){{\footnotesize(a)}}\put(-225,-8){{\footnotesize(b)}} \put(-75,-8){{\footnotesize(c)}}
\caption{(a) The diphoton $\aco$ distribution for events satisfying the signal selection, but before the $\aco < 0.01$ requirement. (b) Diphoton invariant mass and (c) diphoton transverse momentum for events satisfying the signal selection. Data (points) are compared with the sum of signal and background expectations (histograms). Systematic uncertainties of the signal and background processes, excluding that of the luminosity, are shown as shaded bands.}
\label{fig:LbL-control_opt}
\end{center}
\end{figure}

The $\aco\,<\,0.01$ requirement significantly reduces the CEP $gg\rightarrow\gamma\gamma$ background.
Its remaining contribution is evaluated from a control region defined by applying the same selection
as for the signal region, but inverting the $\aco$ requirement to $\aco > 0.01$ (see
Figure~\ref{fig:LbL-control_opt}a), and correcting the measured event yield for the expected signal and $\ggee$ contributions.
The CEP and \ggee\ processes exhibit a significantly broader $\aco$ distribution than the
$\gamma\gamma\rightarrow\gamma\gamma$ process.
In the CEP process gluons recoil against  the Pb nucleus, which  then dissociates.
The shape of the $\aco$ distribution for  \ggee\  events is mainly due to the curvature of the trajectory of the electrons in the detector magnetic field, before they emit hard photons in their interactions with the ID material.

The estimated uncertainty in the CEP $gg\rightarrow\gamma\gamma$ background takes into account the
statistical uncertainty of the number of events
in the $\aco > 0.01$ control region ($17\%$), as well as experimental and modeling uncertainties.
It is found that all experimental uncertainties have negligible impact on the normalization of the CEP $gg\rightarrow\gamma\gamma$ background. The impact of the MC modeling of the $\aco$ shape is estimated using an alternative SuperChic MC sample with no absorptive effects~\cite{Harland-Lang:2015cta}. These effects reflect the absence of secondary particle emissions, which can take place in addition to the $gg\rightarrow\gamma\gamma$ process.
After applying the data-driven normalization procedure, this leads to a $25\%$ change in the CEP background yield in the signal region, which is taken as a systematic uncertainty.
An additional check is done by varying the gluon parton distribution function~(PDF). The differences between the MMHT 2014~\cite{Harland-Lang:2014zoa}, CT14~\cite{Dulat:2015mca} and NNPDF3.1~\cite{Ball:2017nwa} PDF sets have negligible impact on the shape of the CEP diphoton $\aco$ distribution. The background due to the CEP process in the signal region is estimated to be $4\pm1$ events.
In addition, the energy deposition in the ZDC, which is sensitive to dissociation of Pb nuclei,  is studied for events before the $\aco$ selection is imposed.
Good agreement is observed between the normalized CEP expectation from MC simulation and the observed events with a signal corresponding to at least one neutron in the ZDC.

The background contribution from \ggqq\ production is estimated using MC simulation based on \textsc{Herwig}++ and is found to be negligible.
Exclusive two-meson production can be a potential source of background for light-by-light scattering events, mainly due to their similar back-to-back topology. Mesons can fake photons either by their intermediate decay into photons (neutral mesons: $\pi^0$, $\eta$, $\eta'$) or by misreconstructed charged-particle tracks (charged mesons: for example $\pi^+, \pi^-$ states).
Estimates for such contributions are reported in Refs. \cite{KlusekGawenda:2011ib, Klusek-Gawenda:2019ijn,Enterria:2013yra, HarlandLang:2011qd, Harland-Lang:2013ncy} and these contributions are considered to be negligible in the signal region.

The background from other fake diphoton events (mainly those induced by cosmic-ray muons) is estimated using a control region with at least one track reconstructed in the muon system and further studied using the reconstructed photon-cluster time distribution. After imposing the $\pt^{\gamma\gamma}$ requirements, this background  is found to be negligible.
Background from the $\gamma\gamma\rightarrow e^+e^-\gamma\gamma$ reaction is evaluated using the \textsc{MadGraph5$\_$aMC$@$NLO} MC generator~\cite{Alwall:2014hca} and the Pb+Pb photon flux from Starlight. This contribution is estimated to be below 1\% of the expected signal and therefore has negligible impact on the results. The contribution from bottomonia production (for example, $\gamma\gamma\rightarrow \eta_b \rightarrow \gamma\gamma$ or $\gamma\textrm{Pb}\rightarrow \Upsilon \rightarrow \gamma\eta_b \rightarrow 3\gamma$) is calculated using parameters from Refs.~\cite{Ebert:2002pp, Segovia:2016xqb} and considered to be negligible.
The contribution from UPC events where both nuclei emit a bremsstrahlung photon is estimated using calculations from Ref.~\cite{Bertulani:1987tz}. The cross section for single-bremsstrahlung photon production from a Pb ion in the fiducial region of the measurement is calculated to be below \SI{e-4}{\pb} so that the coincidence of two such occurrences is considered to be negligible.

After applying the signal selection, \Nobs events are observed in the data where $\NsigExp\pm4~\textrm{(syst.)}$ signal events and $\Nbkg \pm 1~\textrm{(stat.)} \pm 3~\textrm{(syst.)}$ background events are expected.
The probability that the data are compatible with the background-only hypothesis was evaluated in a narrower $0 < \aco < 0.005$ range, which in studies using simulated data was found to be most sensitive.
In this region, 42 events are observed in the data where $25\pm3~\textrm{(syst.)}$ signal events and $6 \pm 1~\textrm{(stat.)} \pm 2~\textrm{(syst.)}$ background events are expected.
The data excess is quantified by calculating the background-only $p$-value using a profile likelihood-ratio test statistic~\cite{Cowan:2010js}, resulting in an observed (expected) statistical significance of 8.2 (6.2) standard deviations.
Photon kinematic distributions for events satisfying all selection criteria
are shown in Figure~\ref{fig:LbL-control_opt}b--c.
A further cross-check of energy deposits in the ZDC for events in the signal region is performed. The activity in the ZDC agrees with the signal-plus-background expectation.
The analysis is also repeated with a lower minimum photon $\eT$ requirement of \SI{2.5}{\GeV}, yielding more signal events but also an increased relative background contribution. Consistent results were found using this relaxed signal selection.
 
The cross section for the $\gamma\gamma\rightarrow \gamma\gamma$ process is measured in a fiducial phase space, defined by a set of requirements on the diphoton final state, reflecting the selection at reconstruction level~\cite{footnote_1}.
Experimentally, the fiducial cross section is given by $\sigmafid\,=\,(N_{\textrm{data}}-N_{\textrm{bkg}})/(C \times \int L d t)$, where
$N_{\textrm{data}}$ is the number of selected events in data,
$N_{\textrm{bkg}}$ is the number of background events,
$\int L d t\,=\,\SI{1.73 \pm 0.07}{\inb}$
is the integrated luminosity of the data sample and $C$ is an overall correction factor that accounts for efficiencies and resolution effects.
The $C$ factor is defined as the ratio of the number of selected MC signal events passing the selection and after applying data/MC correction factors to the number of generated MC signal events
satisfying the fiducial requirements. It is found to be $C = \CFactor \pm \dCFactor$.
The uncertainty in $C$ is estimated by varying the data/MC correction factors within their uncertainties,
as well as using an alternative signal MC sample based on calculations from Ref.~\cite{Klusek-Gawenda:2016euz}.
The probability of additional inelastic interactions in the same bunch crossing is estimated to be 0.3\% and has negligible impact on the signal efficiency.
The overall uncertainty is dominated by uncertainties in the photon reconstruction efficiency (4\%) and the trigger efficiency (2\%).
The uncertainty of the integrated luminosity is derived, following a methodology similar to that detailed in Ref.~\cite{DAPR-2013-01}, from a calibration of the luminosity scale using $x$--$y$ beam-separation scans performed in November 2018.

The measured fiducial cross section is
\SigmaMeas{},
which can be compared with the predicted values of
\SI{45 \pm 5}{\nb} from Ref.~\cite{Enterria:2013yra},
\SI{51 \pm 5}{\nb} from Ref.~\cite{Klusek-Gawenda:2016euz} and
\SI{50 \pm 5}{\nb} from SuperChic 3.0 MC simulation~\cite{Harland-Lang:2018iur}.
The experiment-to-prediction ratios are
$1.73\pm0.40$, $1.53\pm0.33$ and $1.56\pm0.33$, respectively.
 
In summary, this Letter reports the observation of light-by-light scattering in quasi-real photon interactions from ultraperipheral \PbPb\ collisions at \sqrtNN\,=\,\SI{5.02}{\TeV} recorded in 2018 by the ATLAS experiment.
After applying all selection criteria, \Nobs data events are observed in the signal region, while $\Nbkg\pm\dNbkg$ background events are expected. The dominant background processes, i.e. CEP $gg\rightarrow \gamma \gamma$, \ggee\ as well as other fake-photon backgrounds, are estimated from data.
The statistical significance against the background-only hypothesis is found to be 8.2 standard deviations.

\section*{Acknowledgments}
 
We thank CERN for the very successful operation of the LHC, as well as the
support staff from our institutions without whom ATLAS could not be
operated efficiently.
 
We acknowledge the support of ANPCyT, Argentina; YerPhI, Armenia; ARC, Australia; BMWFW and FWF, Austria; ANAS, Azerbaijan; SSTC, Belarus; CNPq and FAPESP, Brazil; NSERC, NRC and CFI, Canada; CERN; CONICYT, Chile; CAS, MOST and NSFC, China; COLCIENCIAS, Colombia; MSMT CR, MPO CR and VSC CR, Czech Republic; DNRF and DNSRC, Denmark; IN2P3-CNRS, CEA-DRF/IRFU, France; SRNSFG, Georgia; BMBF, HGF, and MPG, Germany; GSRT, Greece; RGC, Hong Kong SAR, China; ISF and Benoziyo Center, Israel; INFN, Italy; MEXT and JSPS, Japan; CNRST, Morocco; NWO, Netherlands; RCN, Norway; MNiSW and NCN, Poland; FCT, Portugal; MNE/IFA, Romania; MES of Russia and NRC KI, Russian Federation; JINR; MESTD, Serbia; MSSR, Slovakia; ARRS and MIZ\v{S}, Slovenia; DST/NRF, South Africa; MINECO, Spain; SRC and Wallenberg Foundation, Sweden; SERI, SNSF and Cantons of Bern and Geneva, Switzerland; MOST, Taiwan; TAEK, Turkey; STFC, United Kingdom; DOE and NSF, United States of America. In addition, individual groups and members have received support from BCKDF, CANARIE, CRC and Compute Canada, Canada; COST, ERC, ERDF, Horizon 2020, and Marie Sk{\l}odowska-Curie Actions, European Union; Investissements d' Avenir Labex and Idex, ANR, France; DFG and AvH Foundation, Germany; Herakleitos, Thales and Aristeia programmes co-financed by EU-ESF and the Greek NSRF, Greece; BSF-NSF and GIF, Israel; CERCA Programme Generalitat de Catalunya, Spain; The Royal Society and Leverhulme Trust, United Kingdom.
 
The crucial computing support from all WLCG partners is acknowledged gratefully, in particular from CERN, the ATLAS Tier-1 facilities at TRIUMF (Canada), NDGF (Denmark, Norway, Sweden), CC-IN2P3 (France), KIT/GridKA (Germany), INFN-CNAF (Italy), NL-T1 (Netherlands), PIC (Spain), ASGC (Taiwan), RAL (UK) and BNL (USA), the Tier-2 facilities worldwide and large non-WLCG resource providers. Major contributors of computing resources are listed in Ref.~\cite{ATL-GEN-PUB-2016-002}.
 

\printbibliography
 
\clearpage
 
\begin{flushleft}
{\Large The ATLAS Collaboration}

\bigskip

G.~Aad$^\textrm{\scriptsize 101}$,    
B.~Abbott$^\textrm{\scriptsize 128}$,    
D.C.~Abbott$^\textrm{\scriptsize 102}$,    
O.~Abdinov$^\textrm{\scriptsize 13,*}$,    
A.~Abed~Abud$^\textrm{\scriptsize 70a,70b}$,    
K.~Abeling$^\textrm{\scriptsize 53}$,    
D.K.~Abhayasinghe$^\textrm{\scriptsize 93}$,    
S.H.~Abidi$^\textrm{\scriptsize 167}$,    
O.S.~AbouZeid$^\textrm{\scriptsize 40}$,    
N.L.~Abraham$^\textrm{\scriptsize 156}$
% ...
\bigskip
\\

% ...
$^{13}$Institute of Physics, Azerbaijan Academy of Sciences, Baku; Azerbaijan.\\
% ...
$^{40}$Niels Bohr Institute, University of Copenhagen, Copenhagen; Denmark.\\
% ...
$^{53}$II. Physikalisches Institut, Georg-August-Universit\"{a}t G\"ottingen, G\"ottingen; Germany.\\
% ...
$^{70}$$^{(a)}$INFN Sezione di Pavia;$^{(b)}$Dipartimento di Fisica, Universit\`a di Pavia, Pavia; Italy.\\
% ...
$^{93}$Department of Physics, Royal Holloway University of London, Egham; United Kingdom.\\
% ...
$^{101}$CPPM, Aix-Marseille Universit\'e, CNRS/IN2P3, Marseille; France.\\
$^{102}$Department of Physics, University of Massachusetts, Amherst MA; United States of America.\\
% ...
$^{128}$Homer L. Dodge Department of Physics and Astronomy, University of Oklahoma, Norman OK; United States of America.\\
% ...
$^{156}$Department of Physics and Astronomy, University of Sussex, Brighton; United Kingdom.\\
% ...
$^{167}$Department of Physics, University of Toronto, Toronto ON; Canada.\\
% ...
$^{a}$ Also at Centre for High Performance Computing, CSIR Campus, Rosebank, Cape Town; South Africa.\\
$^{b}$ Also at CERN, Geneva; Switzerland.\\
% ...
$^{*}$ Deceased

\end{flushleft}


\end{document}